# RESILIENT ICT4D: BUILDING AND SUSTAINING OUR COMMUNITY IN PANDEMIC TIMES

Silvia Masiero, University of Oslo, Norway, silvima@ifi.uio.no

Petter Nielsen, University of Oslo, Norway, pnielsen@ifi.uio.no

The impacts of the COVID-19 pandemic, disproportionally affecting vulnerable people and deepening pre-existing inequalities (Drèze, 2020; Qureshi, 2021), have interested the very same "development" processes that the IFIP Working Group 9.4 on the Implications of Information and Digital Technologies for Development has dealt with over time. A *global* development paradigm (Oldekop et al., 2020) has emerged in response to the global nature of the crisis, infusing new meaning in the spirit of "making a better world" with ICTs (Walsham, 2012) that always have characterised ICT4D research. Such a new meaning contextualises our research in the landscape of the first pandemic of the datafied society (Milan & Trerè, 2020), coming to terms with the silencing of narratives from the margins within the pandemic (Milan et al., 2021) – in Qureshi's (2021) words, a "pandemics within the pandemic" producing new socio-economic inequities in a state of global emergency.

The landscape of the pandemic has affected the way ICT4D research is conducted, discussed and communicated. The conduct of "fieldwork" as we knew it in a world of few mobility restrictions, as well as the community participation that has systematically characterised ICT4D research production over the last decades (Walsham, 2017) have become impracticable. In a world of movement restrictions, the "new normal" are digital interactions. Conferencing, the space where research is communicated and our community is built and rebuilt, constituting the soul of IFIP 9.4, has had to be equally rebuilt for a digital world, changing the taken-for-granted practices that have characterised the community for the past decades. Resilience, seen with Heeks and Ospina (2019) as "the ability of systems to cope with external shocks and trends", has become fundamental in the making of research and in the interactive aspects of communicating it.

It was these considerations that brought us to choose *Resilient ICT4D* as the theme of the 1st IFIP 9.4 Virtual Conference on Implications of Information and Digital Technologies for Development, which took place on 26-28 May 2021. The Conference was, in the first place, an occasion to leverage the digital means to abate the barriers – of geography, geopolitical constraints, movement costs and viability – that characterise conferencing in the physical world, and whose removal has allowed community conversations that would have been very difficult in the physical world scenario. Organised virtually, with a shared governance model led by a collective of 32 Track Chairs without a host institution, the event was also a route to questioning the unilateral organisational model taken for granted when conferencing. Such a combination of the digital means with shared governance has afforded the creation of 13 conference tracks, reflecting communities from different walks of ICT4D and generating new occasions for cross-community mutual learning.

## A DIFFERENT PHD DAY
When launching the First IFIP 9.4 Virtual Conference, the question of how to organise a PhD Day – a space for PhD candidates from around the globe to convene and discuss research and the academic world – immediately came to mind. One route to do this, following the blueprint of multiple "doctoral consortia" from academic disciplines, was that of formulating a call for applications, then selecting the "best" ones and forming a small group of 15-20 to run the event. Such a model would have had some value: small groups, it is known, allow for better interaction and greater room for one-to-one mentoring than large ones. They also afford discussion of people's specific research topics, often ending up focusing on the "publishability" of the material and various strategies to "thrive" as productive researchers in the academic world.

Leveraging the virtual means, the PhD day at the IFIP 9.4 Virtual Conference chose to propose an alternative to such a paradigm. We started with the choice of uncapped participation: every PhD candidate who registered, with a deadline of 11 May, was invited to join the event and the discussions taking place in it. This is how a group of 74 PhD candidates, from 27 countries and mentored by 25 faculty members, was formed, and participated in three parallel sessions – conveniently named "networking-mentoring tables" – throughout the day. The online means, creating the possibility to divide the group into parallel rooms and putting together students and faculty, afforded the possibility to create as many groups as requested, fostering meaningful conversations and bypassing the idea of "selecting the best applications" entrenched in many doctoral consortia.

Significant was, in a similar vein, the choice of the panels held during the Conference PhD day. The first panel, named "Lessons from the PhD Journey", was conveyed by five colleagues – two close to finishing their PhDs, three who finished recently – to share insights from the PhD journey with the group, fostering many questions especially from candidates that made a more recent start into their doctoral programmes. The second panel, named "Academic Careers with a Human Face", brought together four academics reflecting on the challenges that an academic career brings to us: reflections covered the (lack of) sustainability of an academic sector that favours hyperproductivity over mental and physical health, and strategies were shared, by all four panellists, to imagine and live academia in ways that respect the work-life balance and enhance happiness. Participation was strong, and both panels acted in constructive, but open, contrast with the "publishing panels" typical of academic conferences, where the importance of horizontal learning and the preservation of health and happiness are less than frequently noted and emphasised.

## A SHARED GOVERNANCE MODEL

While conferencing in the physical world involves – by its very nature – the presence of a host institution, taking care of logistics and organisation, the same does not hold for virtual conferencing. As opposed to events held in a physical space, led and managed by one (or more) institutions related to it, online conferencing allows something different: a model where the "host" is indeed the virtual space, open and accessible to everyone pending infrastructural constraints. The most evident barrier this bypasses – participation of colleagues who would otherwise be constrained by financial, travel-related or visa restrictions – is accompanied by a more hidden barrier, that of unilateral decision making tied to location and organisational responsibilities. It is this second barrier that a shared governance model, independent of institutions or constraints, aims to overcome.

The First IFIP 9.4 Virtual Conference was announced in December 2020, with a Call for Tracks that invited colleagues from inside – and, crucially, outside – ICT4D to propose conference tracks, with the note that all Track Chairs would be conference organisers. Having received 14 track proposals, bringing together 32 Track Chairs from different backgrounds, fields and walks of ICT4D, the Conference was run by the Track Chairs committee as a collective that, meeting periodically in the run-up to the event, took crucial decisions on tracks, keynote speaker invitations, proposed panels and workshops, important practical matters and topics to be engaged by the group. Such a varied committee was then responsible for the programme and is making: advertising the event through multiple networks, building on such netwoks' knowledge and community expertise, afforded reaching out beyond the core that characterised the tradition of IFIP 9.4 for the last decades. Of the 13 Conference Tracks featured in the event, some appeared this year for the first time: among others, ICTs and Data Justice, Feminist and Queer Approaches to Information Systems in Developing

Countries, and Potential and Risks of ICTs for Development. These opened new terrains of discussion in an established conference group.

Arisen as a result of the affordance of digital means, the shared governance model proposed at the Conference allowed us to discover, and indeed leverage, a way of conferencing that distributes decision-making across a collective, maximising the representation of participating groups and affording conversation on key decisional aspects. Reflecting *ex-post* on such an experience, we find great value in a model that makes voicing distributed rather than unilateral, and believe that such a model – with the due changes and adaptations – can be made to persist in a post-pandemic physical world, leveraging the lessons learned here towards the ubiquity of community voicing.

## FREE AND OPEN ACCESS

The 1st virtual IFIP 9.4 conference registration was 0 (in any currency), and participants could register until the last day of the Conference. The committee edited and published the proceedings on the IFIP 9.4 website. Thus, in a spirit of openness, anyone could participate at the Conference and the papers are not behind payment walls but openly available for anyone to access and read.

## ACROSS TOPICS, ACROSS FIELDS

Resulting from the above-mentioned Call for Tracks, the First IFIP 9.4 Virtual Conference grouped together a wide span of topics from within and beyond the core of ICT4D. The 13 Tracks featured in the Conference were:

- ICT and Resilience Building: Climate Change, Pandemic, and Other Stressors
- Digital Platforms in, from and in-between the Global South and North
- Data Science in Public Health
- ICT4D and Data Justice
- Our Digital Lives (IFIP 9.5 Track)
- Digital Social Enterprises & COVID-19: Enablers, Sustainability & Pathways
- Feminist and Queer Approaches to Information Systems in Developing Countries
- Displacements, ICTs, and #NewNormal
- Digital Authoritarianism and Fundamentalism: Problems and Solutions
- The Role of ICT in Achieving Social Justice (ICT4SJ)
- Potential and risks of advanced technologies in the Global South
- Digitalisation for Indigenous Emancipation
- General Track

Beyond the convening of tracks held this year for the first time at IFIP 9.4, the Conference featured Track 5, "Our Digital Lives", managed and convened by IFIP WG 9.5 and bringing together papers that explored the intersections between the two IFIP Working Groups. Track 5 was a great experience of collaboration between IFIP WGs, and we look forward to building on the interactions created there to work towards future collaborative events.

Spanning across fields were also the four panels of practitioners and colleagues that the Conference featured. The panels convened in the event were respectively on:

- Open Data in the Global South: Challenges and Lessons Learned
- Digital Labour in the Global South
- Deconstructing Notions of Resilience
- Feminist Approaches to Information Systems and Digital Technologies for Development

All panels, chaired by members of the IFIP 9.4 Community, featured colleagues from research, practice and civil society who brought insights on the immediate relevance of the panel's topics for Resilient ICT4D. We can't help noticing, once again, how panel compositions from across countries, contexts and regions were made possible by digital means that democratised conference participation, allowing exchanges of insights that would have been made impossible or very difficult by physical barriers.

By the same token, what the Conference afforded was encounters of colleagues and potential authors with the Editors of three Special Issues, launched by journals from the discipline and holding currently open, or soon to be opened, Calls for Papers:

- Information Technology for Development Special Issue on Understanding Local Social Processes in ICT4D research
- MIS Quarterly Special Issue on Social Justice
- Information Systems Journal & Electronic Journal of Information Systems in Developing Countries Special Issue on Digital Transformation in Latin America: Challenges and Opportunities.

In the Special Issue workshops run for each Call for Papers, Editors established a conversation with the audience and colleagues with paper ideas, creating spaces of reflection on how to bring such ideas forward. It should be noted that the format was not created in open opposition with the "Meet the Editors" panels that are common in mainstream discipline conferences, usually characterised by crammed rooms and a hierarchical dichotomy of an audience of early-career researchers striving for tenure – and a panel of seniors dictating directions for publishing, with varying degrees of empathetic understanding of the struggles lived by early-career colleagues. However, the spaces of horizontal conversation created through the virtual space served as a route to circumvent that dichotomy, enabling author-editor conversations and problematising the idea of "learn how to publish in a top journal" that is somehow transmitted. As one of the Editors who joined put it, the Conference was an occasion for the editors themselves to get to know the IFIP 9.4 community and start a process of interaction with it.

## WHERE NEXT?

The original reason for a virtual convening of the IFIP 9.4 Conference was the postponing of the 16$^{th}$ edition, planned for Lima in 2021, to the next year, given the COVID-19 pandemic. The current conditions, with the impacts of the pandemic perduring on a global scale, do not make it possible to know the likelihood of an in-person meeting in a year from now. What is visible, however, are the lessons learned in need to arrange an online gathering of over 400 registered people, and the decreased barriers and shared governance model that this experience of "Resilient ICT4D" has entailed.

As a result, we believe the lessons of resilience, solidarity and interactions learned in these pandemic times are here to stay. The hope of in-person meetings in the near future remains strong – what is not lost however, is the learnings that over a year in a digital world has entailed. Looking forward from the experience of our first virtual IFIP 9.4 Conference, what we see is a free, open-access, democratising conference model becoming known to the community, and the many reasons to embed the lessons of such a model in the conferences and events that will come next.